\documentclass[a4paper,11pt]{article}
\pdfoutput=1 

\usepackage[T1]{fontenc} 
\usepackage{amsfonts,amssymb,amsmath,mathbbol}
\usepackage{graphics,graphicx,epsfig,ulem,color}
\usepackage{subfigure}
\usepackage{cite}
\usepackage{notoccite}

\usepackage{fancyhdr}

\begin{document}
\bibliographystyle{unsrt}

\pagestyle{plain}
\lhead{} 
\rhead{IPPP/14/71, DCPT/14/142} 

\numberwithin{equation}{section}
\title{{\normalsize  \mbox{}\hfill ~}\\
\vspace{2.5cm}
\Large{\textbf{Hidden photons with Kaluza-Klein towers\vspace{0.5cm}}}}

\author{Joerg Jaeckel$^1$, Sabyasachi Roy$^2$, Chris J. Wallace$^2$\\[2ex]
\small{\it $^1$ Institut f\"ur theoretische Physik, Universit\"at Heidelberg, Philosophenweg 16, 69120 Heidelberg, Germany}\\[0.5ex]
\small{\it $^2$ Institute for Particle Physics Phenomenology, University of Durham, South Road, Durham, DH1 3LE, UK}\\[0.5ex]
}

\maketitle
\thispagestyle{fancy}

\begin{abstract}
One of the simplest extensions of the Standard Model (SM) is an extra U(1) gauge group under which SM matter does not carry any charge. The associated boson -- the hidden photon -- then interacts via kinetic mixing with the ordinary photon.
Such hidden photons arise naturally in UV extensions such as string theory, often accompanied by the presence of extra spatial dimensions. In this note we investigate a toy scenario where the hidden photon extends into these extra dimensions.
Interaction via kinetic mixing is observable only if the hidden photon is massive. In four dimensions this mass needs to be generated via a Higgs or Stueckelberg mechanism. However, in a situation with compactified extra dimensions there automatically exist massive Kaluza-Klein modes which make the interaction observable.
We present phenomenological constraints for our toy model.
This example demonstrates that the additional particles arising in a more complete theory can have significant effects on the phenomenology.
\end{abstract}

\section{Introduction}
Extra U(1) gauge groups and the corresponding gauge bosons are a common feature in completions of the Standard Model (SM)~\cite{Okun:1982xi,Galison:1983pa,Holdom:1985ag,Dienes:1996zr,Lukas:1999nh,Abel:2003ue,Blumenhagen:2005ga,Abel:2006qt,Abel:2008ai,Goodsell:2009pi,Goodsell:2009xc,Goodsell:2010ie,Heckman:2010fh,Bullimore:2010aj,Cicoli:2011yh,Goodsell:2011wn,Ahlers:2008qc} such as, e.g. string theory.
To account for their not having being observed these gauge bosons either have to be very massive or the interactions with SM particles have to be very weak. Here, we are mainly interested in the latter case.

Weakness of the interactions is automatic if SM particles do not carry charge under this new U(1). In this case the
only renormalizable interaction is via kinetic mixing of the new U(1) with the photon\footnote{In principle the interaction is via the hypercharge U(1) but after electroweak symmetry breaking the most relevant part for the accessible phenomenology at low energies is the mixing with the photon.},
\begin{equation}
\label{simplelag}
{\mathcal L} =-\frac{1}{4} F^{\mu\nu}F_{\mu\nu}-\frac{1}{4}X^{\mu\nu}X_{\mu\nu}-\frac{1}{2}\chi F^{\mu\nu}X_{\mu\nu}
+j^{\mu}A_{\mu}.
\end{equation}
Here $F^{\mu\nu}$ is the ordinary photon field strength tensor, $X^{\mu\nu}$ is the field strength for the hidden photon field $X^{\mu}$ and $j^{\mu}$ is the electromagnetic current.
In the simplest case where there is no additional matter we do not have such a current interaction with the hidden photon since all SM particles are uncharged under it.

It is straightforward to see that if the Lagrangian above was a complete description of the new gauge boson, no effects would be observable, since the kinetic mixing term can be made to vanish by applying a field redefinition,
\begin{equation}
X_{\mu} \to X_{\mu}-\chi A_{\mu}.
\end{equation}
In the presence of a mass term,
\begin{equation}
{\mathcal L}_{\rm mass}=\frac{1}{2} m^{2}_{X} X^{\mu}X_{\mu},
\end{equation}
this redefinition does not remove all couplings between $A$ and $X$ and the hidden photon is therefore observable.

In four dimensions such a mass term has to be generated by a Higgs or Stueckelberg mechanism.
Interestingly, if we have extra compactified spatial dimensions, which are commonly present in string theory,
there are automatically massive Kaluza-Klein (KK) modes.
If the hidden photon extends into the extra dimensions these modes (and their kinetic mixing with the photon) can facilitate an observable interaction with the SM. We will explicitly show this for a toy model in Sect.~\ref{sec:model}.
In this case there is no need for an additional mass generation mechanism and we focus on this situation.

At the same time the presence of a potentially huge number of observable modes impacts the phenomenology and can impose strong constraints (this may also open new paths for experimental or observational tests).
To make a start on this we examine the existing constraints on the parameter space of a single four dimensional hidden photon and investigate their impact in our toy model featuring a KK tower.

We will focus on the case of large, flat extra dimensions (LED)~\cite{ArkaniHamed:1998rs,ArkaniHamed:1998nn}.
The case of a warped, Randall-Sundrum \cite{Randall:1998uk,Randall:1999ee,Randall:1999vf} type scenario has been considered in~\cite{McDonald:2010iq,McDonald:2010fe}.

The paper is structured as follows. In Section~\ref{sec:model}, we present our toy model for a higher dimensional hidden photon field in a braneworld scenario, in the language of effective field theory. In Section~\ref{sec:4Dbounds}, we briefly review the existing constraints on 4-dimensional hidden photons, and discuss in general terms features and their relevance for higher dimensional scenarios.  We analyze their quantitative impact  in Section~\ref{sec:results}. Further discussion of some issues regarding renormalizability and perturbativity is given in Section~\ref{sec:pert}. Finally, we
conclude in Section~\ref{sec:conc}.

\section{Toy Model}
\label{sec:model}

We begin with a $D$-dimensional bulk of Minkowski space, with the Standard Model confined to a 3-brane within the bulk. Our description is purely a low-energy, effective one; we assume the brane to be infinitely heavy, infinitesimally thin and possessed of no intrinsic dynamics. We allow gravity and an additional U(1)$'$ gauge group, which gives rise to a hidden photon, to propagate freely though the whole bulk.

The interaction between the photon and the new ``hidden photon'' is introduced through a brane localized mixing term.
This is not the most general situation because we have neglected a number of brane localized kinetic and interaction terms. Nevertheless this simplified toy model serves to demonstrate the potentially huge impact the KK tower can have on phenomenology.

The effective action describing the above setup then is:
\begin{equation}
S_{D} = \int \textrm{d}^D x \sqrt{g} \bigg( -\frac{1}{4}F^{\mu\nu}F_{\mu\nu}\delta^{d}(\vec{y}) -\frac{1}{4}X^{MN}X_{MN}
-\frac{1}{2} \chi_{\rm D} F^{\mu\nu}X_{\mu\nu}\delta^{d}(\vec{y})\bigg),
\label{eq:hdaction}
\end{equation}
The captial roman indices $(M,N)$ run over $0,\dots,D-1$, the lower case greek indices $(\mu,\nu)$ over $0,\dots,3$ and the coordinate $\vec{y}$ represents $\{x^{4},\dots,x^{D-1}\}$.

We compactify the $d=D-4$ additional spatial dimensions from Minkowski space onto a $d$-dimensional torus $\mathcal{T}_{d} = \mathcal{S}_1 \times \mathcal{S}_1 \dots \mathcal{S}_1$, with each torus chosen to be of equal radius $R$. In this case the 4-dimensional Planck mass, $M_{\rm Pl}$, is related to the higher dimensional one, $M_{\star}$ via,
\begin{equation}
M^{2}_{\rm Pl}= M^{2+d}_{\star} (2\pi R)^{d}.
\label{eq:planck}
\end{equation}
Limits on $R$ arise both from direct tests of the gravitational inverse square law and via limits on $M_{\star}$ from graviton searches at colliders. They are summarized in Table~\ref{tab:limits}.
%
\begin{table}
\centerline{\begin{tabular}{| c | c | c |}
\hline
$d$ & $1/R=m_0$ & $M_{\star}$\\
\hline
1 & $>200$~$\mu$eV & $\gtrsim3\times10^5$~TeV \\
2 & $>700$~$\mu$eV & $\gtrsim3$~TeV \\
3 & $>100$~eV & $\gtrsim3$~TeV \\
4 & $>50$~keV & $\gtrsim3$~TeV \\
5 & $>2$~MeV & $\gtrsim3$~TeV \\
6 & $>20$~MeV & $\gtrsim3$~TeV \\
\hline
\end{tabular}}
\caption[]{
Limits on the scale of extra dimensions. The $d=1$ constraint arises from direct tests of the gravitational inverse square law~\cite{Adelberger:2003zx,Hoyle:2004cw}. For $d=$2--6 the scale is more strongly constrained by the minimum value of the extra dimensional Planck scale $M_{\star}$~\cite{Aad:2012cy,Chatrchyan:2012me}, provided by colliders.}
\label{tab:limits}
\end{table}

For the hidden photon, which lives in the bulk, compactification results in a stack of Kaluza-Klein modes corresponding to the Fourier modes of the hidden photon field in the extra spatial dimensions.
In the general case the KK mode expansion is,
\begin{equation}
X_{M}(x^{\mu},y^{a}) = \frac{1}{(2\pi R)^{d/2}} \sum_{|\vec{n}|\ge 0,\sigma} X_{M}^{(\vec{n},\sigma)}(x^{\mu}) \prod_{i=1}^{d} f_{\sigma_{i}}(n_{i} y_{i})
\end{equation}
where $\sigma = \{(+,\dots,+,+),(+,\dots,+,-), \dots, (-,\dots,-,-)\}$ and,
\begin{equation}
f_{s}(n x) = \left\{
\begin{array}{cccc}
\sqrt{2}\cos{n x} & \quad & s = +, & n > 0\\
\sqrt{2}\sin{n x} & \quad & s = -, & n \geq 0\\
1 & \quad & s = +,& n = 0.\\
\end{array}
\right.
\end{equation}
For example, in the simplest case of a single extra spatial dimension ($d=1$) we have,
\begin{equation}
X_{M}(x^{\mu},y^{a}) = \frac{X_{M}^{(0)}}{(2 \pi R)^{1/2}} + \frac{1}{(\pi R)^{1/2}} \sum_{n>0} \left( X^{(n,+)}_{M}(x^{\mu})\cos{\left(\frac{n y}{R}\right)}
+ X^{(n,-)}_{M}(x^{\mu})\sin{\left(\frac{n y}{R}\right)}\right).
\end{equation}

By choice, our SM 3-brane is located at $\vec{y}=0$. Accordingly, the kinetic mixing operator is also localized at $\vec{y}=0$. Therefore only those hidden photon modes that are non-vanishing at this point can interact with the SM photon. In the expansion for the 5D case, this corresponds to the cosine mode. In the general case all $f_{s}$ factors must be cosines corresponding to $\sigma = (+,\dots,+)$. (Of course, the brane may be located anywhere in the bulk, and shifting it would result in a linear combination of the $X^{+}$ and $X^{-}$ fields interacting, which could be simplified by a field redefinition). The zero mode has no mass, hence its mixing is not observable. Inserting the Fourier transform into the action, we obtain for the $D$-dimensional case:
\begin{align}
S_{\textrm{eff}} = \int d^4 x\,\bigg[ &-\frac{1}{4}F^{\mu\nu}F_{\mu\nu} -\frac{1}{4}X^{\mu\nu(0)}X_{\mu\nu}^{(0)}
+ \sum_{|\vec{n}|\ge 0,\sigma} \bigg(\frac{1}{4}X^{\mu\nu(\vec{n},\sigma)}X_{\mu\nu}^{(\vec{n},\sigma)} +  \frac{1}{2}\frac{n^2}{R^2}X_{\mu}^{(\vec{n},\sigma)}X^{\mu\,(\vec{n},\sigma)}\bigg)
\nonumber\\
&+ \sum_{|\vec{n}|\ge 0} \bigg(\frac{1}{2} \chi_{4} F^{\mu\nu}X_{\mu\nu}^{(\vec{n},+\ldots+)} \bigg) 
+\ldots\bigg]
\label{eq:action}
\end{align}
Omitted from the Lagrangian are the $d$ massless scalar fields from $X^{0}_{a=5\dots D}$, and the Kaluza-Klein tower associated with $d-1$ (linear combinations) of the scalars. The modes of the remaining scalar Kaluza-Klein tower are eaten by the respective massless vector in the KK stack, generating a single stack of 4D hidden photons, with masses
\begin{equation}
m^2_{\gamma'} = \frac{n^2}{R^2}(1+{\mathcal{O}}(\chi^2)).
\end{equation}
In the  action~\eqref{eq:hdaction}, $X_{M}$ is a $D$-dimensional field with mass dimension $\frac{D}{2}-1$. When we write the 4-dimensional effective action~\eqref{eq:action}, the mass dimension for the KK modes is reduced to 1. The necessary rescaling of the hidden photon field by a volume factor results in the same rescaling of the kinetic mixing parameter (generalizing to the $d$-dimensional case),
\begin{equation}
\chi_4 = \frac{\chi_{D}}{(\pi R)^{d/2}} \prod_{i=1}^{d}\eta_{n_{i}},
\qquad\textrm{where}\qquad
\eta_{n_{i}} = \left\{
\begin{array}{ccc}
1/\sqrt{2} & \quad & n_{i} = 0\\
1 & \quad & n_{i} \neq 0\\
\end{array}
\right. ,
\label{eq:chi}
\end{equation}
where $n_{i}$ is the component of $\vec{n}$ along the $i^{\rm th}$ extra dimension.
The sums in Eq.~\eqref{eq:action} run over $\vec{n} \ge 0$, representing the fact that (for $d>1$) massive KK modes arise even if momentum along some of the extra dimensions is zero.
In practice, for the scenarios we consider there are very many modes, and few have a zero momentum component compared to the number that do not. For phenomenological purposes, we therefore ignore the product in Eq.~\eqref{eq:chi}, and use the same equation, $\chi_4 \sim \chi_D / (2 \pi)^{d/2}$, for all modes.

\subsection{The particle spectrum}

After compacitifcation, the resulting particle spectrum consists of the unmodified SM field quanta, which interact with the hidden sector only through kinetic mixing with a stack of infinitely many additional hidden photons. The mass of the lightest of the hidden photons, $m_0$, is determined by the compactification radius, and each subsequent field is $m_0$ more massive than the last. We have a stack of increasingly massive Kaluza-Klein modes, corresponding to the infinite number of terms in a Fourier decomposition of the original, higher dimensional field along the additional dimensions. We also have a similar stack of scalar fields associated with each extra dimension (one of which is eaten), which decouple from our phenomenology.

\subsection{Gravity}

In a scenario featuring large extra dimensions the fundamental Planck scale can be significantly lowered (see Eq.~\eqref{eq:planck}) and the effects of gravity in the bulk can be important.

The interaction of gravity with any theory described by an energy momentum tensor $T^{\mu\nu}$ is,
\begin{equation}
	\mathcal{L} = -\frac{1}{2 M_{\rm Pl}}\sum_{\vec{k}}\left(G^{(\vec{k})}_{\mu\nu} T^{\mu\nu} + \sqrt{\frac{2}{3(d+2)}}\phi^{(\vec{k})}T^{\mu}_{\mu} \right).
\end{equation}
It is clear than any interaction involving the graviton $G_{\mu\nu}$ or dilaton $\phi$ is suppressed by the Planck scale. For a compactified theory, the fundamental scale is reduced after a summation over all modes (i.e. when the extra dimension is resolved), but each interaction is still suppressed by $M_{\rm Pl}$. Since both the gravitational and U(1)$'$ field begin as massless and are compactified on the same manifold, their mass spectra are the same, with each mode having mass $|k|/R$.  Thus the same number of graviton and hidden photon modes contribute to a given process, so the analysis for one mode is true in generality, and pure gravitational interactions are important relative to hidden photon interactions only when $\chi_{4} \lesssim \frac{\Lambda}{M_{\rm Pl.}}$, where $\Lambda$ represents the energy scale at which the relevant process occurs. As we will see, such small values of $\chi_{4}$ are out of reach of current experiments -- even with a stack of KK modes present -- and as such we may safely ignore gravitational interactions. We discuss hidden photon decays into gravitons below.

\subsection{Bulk Interactions}

For interactions between gravity and the hidden photon, i.e. those involving only bulk fields, one may ignore the presence of the 3-brane and there is translational invariance along the extra dimensions in the bulk -- a symmetry group (generated by the position operator) of which the conserved quantity is momentum. Upon compactification, the extra dimensional momentum conservation translates into conservation of KK mode number. This forbids decays of KK particles of the form $X^{(k)} \to X^{(l)} + G^{(m)}$, where all the products are bulk fields, and $k$, $l$ and $m$ are the KK mode numbers. Conservation of mode number is encapsulated in the relation $|k| \leq |l + m|$ and, since any decay can only be to lighter states (recall that the mass of a KK mode is $m = n \times m_0$, with $n$ the mode number), in the best case the phase space vanishes precisely (at $|k| = |l + m|$). Note however, that this is a special feature of the symmetry structure of our toy model.

\subsection{Brane interactions}

The above reasoning holds for interactions occurring purely in the bulk. However, presence of the SM 3-brane breaks the translational invariance so that any interaction involving a brane localized (i.e. SM) particle does not conserve KK number, making decays of the form $X^{(k)} \to G^{(l)} + \gamma$ possible. Nonetheless, the decay is strongly suppressed by the weakness of the gravitational interaction (which is suppressed by the higher dimensional Planck scale $M_{\star}$), and decays of the hidden photon to pure SM states $e^{+}e^{-}$ ($m_{X}>2 m_{e}$), $\gamma\gamma\gamma$ ($m_{X}<2m_{e}$) dominate. The decay to one photon and one graviton is bounded by the case where both decay products are massless, with decay rate,
\begin{equation}
  \Gamma(X^{(k)} \to G^{(l)} + \gamma) < \frac{\chi^2 m_{X}^3}{12 \pi M_{\rm Pl.}^2}.
\end{equation}
After summing over KK modes and using relation \eqref{eq:planck},
\begin{align}
  \Gamma(X^{(k)} \to G + \gamma)_{\rm total} &<
  \Gamma(X^{(k)} \to G^{(0)} + \gamma)\times
  \int_{0}^{m_{X}/m_{0}} d^{d}l\nonumber\\
  &< \frac{1}{2^{d-1} \pi^{d/2} d\,\Gamma(d/2)}
  \frac{\chi^2 m_{X}^{d+3}}{12 \pi M_{\star}^{d+2}}
\end{align}
The case for which decays to $G + \gamma$ are most likely to be relevant is when $d=2$, which minimises the suppression of the branching ratio with $d$, combined with the extra dimensional Planck scale $M_{\star}^{d=2}$ being $10^{-5}$~ times the $d=1$ case.
Inserting the most extreme value, where $m_{X}\sim M_{\star}$ for $m_0$ very low, the decay rate into KK gravitons plus photons is smaller than, but of the same order as $\Gamma(X\to e^+e^-)$ for $m_{X}>2m_{e}$.
This does not affect the LHC constraints we compute, but could potentially lead to new, novel collider signatures and be suitable for investigation in a more developed (i.e. non-toy) model.

The constraints for which additional decay channels to photons are important (for example the Intergalactic Diffuse Photon Background (IDPB) constraint), all occur at energies where the decay rate to gravitons is negligible.

It is pertinent here to mention that the action~\eqref{eq:hdaction} does not include a brane-localized hidden photon term $\propto -\frac{1}{4}X^{\mu\nu}X_{\mu\nu}\delta^{d}(\vec{y})$, which could lead to a different phenomenology owing to additional mixing terms in the effective Lagrangian.

\section{Canonical constraints on the hidden photon parameter space}
\label{sec:4Dbounds}

Since we will be investigating the whole of the currently accessible hidden photon ($m_{X},\chi$) parameter space in a new context, it is worthwhile to summarize the existing limits. Here we provide a list of the ``canonical'' constraints for the 4D case (see, e.g.,~\cite{Jaeckel:2010ni,Jaeckel:2013ija}), where the hidden photon mass is generated by a Higgs or Stueckelberg mechanism. The following bounds involve processes where the hidden photon is produced on-shell, listed in approximate order of increasing energy.

\begin{itemize}
\item{{\bf Cosmic microwave background.} Hidden photons contribute a positive effective neutrino number, causing potential conflict with the number, $N_{\nu}^{\textrm{eff}} = 3.36^{+0.68}_{-0.64}$ \cite{Ade:2013zuv}, obtained using measurements of the cosmic microwave background (CMB), large scale structure (LSS) and supernovae~\cite{Jaeckel:2008fi,Redondo:2010qc}. The low mass region bounded by CMB considerations renders it unimportant for the LED model we employ~\footnote{During the final stages of the preparation of this manuscript, some constraints arising from consideration of the CMB and BBN for higher mass hidden photons with extremely weak couplings (``very dark photons'') were released \cite{Fradette:2014sza}. It would be interesting to consider their impact on the LED scenario in future work.}.}
\item{{\bf Big bang nucleosynthesis.} A small area of parameter space may be ruled out by considering the hidden photon distortion to big bang nucleosynthesis (BBN) \cite{Kolb:1981cx,Redondo:2008ec}. This bound applies only to hidden photons of extremely low mass and is eclipsed by more constraining bounds in our scenario~\footnote{Higher mass constraints were recently detailed in \cite{Fradette:2014sza}, see previous footnote.}.}
\item{{\bf Light-shining-through-walls.} Photon-hidden photon oscillations allow photons to pass through otherwise infinite potential barriers by converting to a hidden photon state that does not interact with the barrier. Lab experiments searching for this ``light-shining-through-walls'' (LSW) phenomena put constraints on the parameter space. For a review, see~\cite{Redondo:2010dp}. More recent developments are reported in~\cite{Betz:2012ce,Dobrich:2013mja,Dobrich:2012jd}. A lower mass regime was recently probed with a few GHz microwave cavity resonator operating on the same principle as optical LSW experiments \cite{Jaeckel:2007ch,Parker:2013fxa,Betz:2013dza,Graham:2014sha}. In our scenario, these constraints are eclipsed by others.}
\item{{\bf Helioscopes.} The CERN-Axion-Solar-Telescope (CAST) operates on the LSW principle, with the sun as the light source~\cite{Redondo:2008aa,Gninenko:2008pz,Lakic:2012fg}.}
\item{{\bf Stellar evolution.} Hidden photons contribute to energy loss from the Sun. Reconciling the existence of a hidden photon with the accepted models of stellar evolution bound large areas of parameter space. Identical arguments can be applied to the evolution of Horizontal Branch stars~\cite{Frieman:1987ui,Raffelt:1987yb,Popov:1999,Redondo:2008aa,An:2013yfc,Redondo:2013lna}.}
\item{{\bf Intergalactic-diffuse-photon-background.} It is possible that hidden photons produced in the early universe have survived until the present day. Their decays to SM photons would contribute to the intergalactic-diffuse-photon-background (IDPB)~\cite{Redondo:2008ec}.}
\item{{\bf Electron fixed target experiments.} An electron beam incident on a target emits hidden photons by bremsstrahlung. The hidden photon can then pass through the shielding before decay via SM photon into an $e^+e^-$ pair, which can be detected~\cite{Bjorken:2009mm}.}
\item{{\bf SN1987a.} Hidden photons contribute to the energy loss in supernovae. If this effect is too large, the supernova may quickly cool preventing the observed neutrino burst. This provides a very strong constraint on hidden photon couplings for a large range of masses \cite{Turner:1987by,Raffelt:1987yt,Bjorken:2009mm}.}
\item{{\bf B factories.} The BABAR \cite{Lees:2014xha} and DA$\Phi$NE (with the KLOE-2 detector) \cite{Babusci:2012cr} collaborations produce constraints on the hidden photon parameter space as a byproduct of their $e^+ e^-$ collisions to study $\Upsilon$, $\phi$ and $\eta$ meson decays. The hidden photon may be produce missing energy or multi-lepton signatures in the minimal model, featuring a Higgs$'$ boson \cite{Batell:2009yf}, which is not present in our model. The search for a hidden photon without an associated Higgs$'$ relies upon a peak search\footnote{The alternative is a missing energy signal. Here, in order for the hidden photon to escape the detector, the mixing parameter must be very small, which suppresses production strongly. This is investigated in more detail for the LHC in Section \ref{sec:LHC}.}, which, in the presence of a KK tower, is not viable in the mass region these facilities can probe, since the mass splitting of the modes is smaller than their width. A full treatment would require simulation of the experimental detector, which is outside the scope of this work, since in the KK scenario the resulting limits are unlikely to improve much on the LHC dilepton search we discuss in Section~\ref{sec:LHC}}.
\item{{\bf LHC.} Recently, constraints on hidden photons from the LHC were computed \cite{Jaeckel:2012yz}. We consider two signals - monojets and dileptons, finding that the dilepton constraint is everywhere stronger than the monojet. Note that this result changes if the hidden photon mediates interactions to a hidden sector, investigated comprehensively in \cite{Frandsen:2012rk}.}
\end{itemize}

\begin{figure}[!ht]
\centerline{\includegraphics[width=0.8\textwidth]{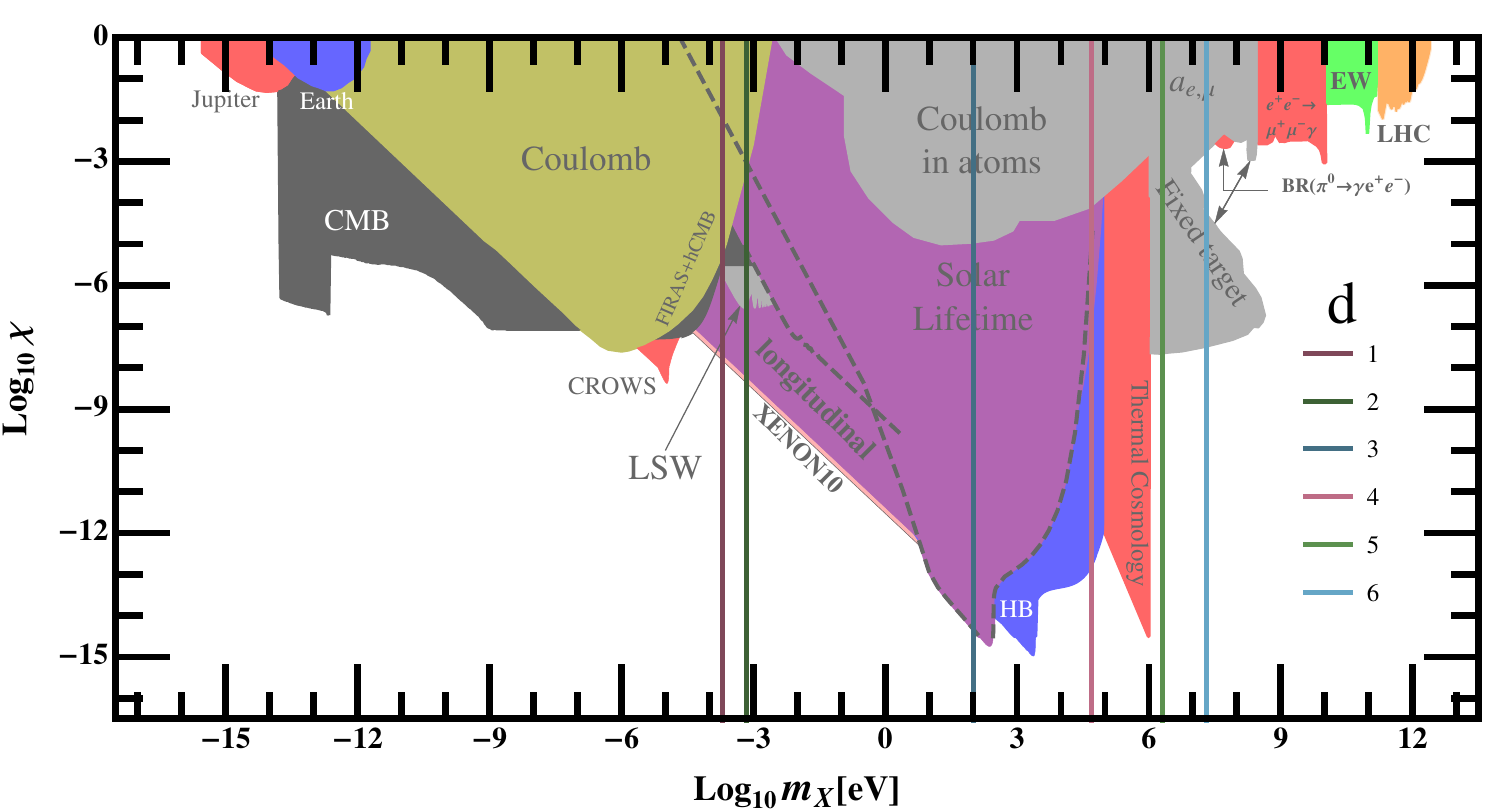}}
\caption[]{Limits on hidden photon parameter space in $D=4$ dimensions. Coloured lines indicate the minimum mode mass associated with the constraint on the compactification radius $R$ for each of $d=1 \dots 6$ extra dimensions.}\label{Fig:4d}
\end{figure}

Next we have bounds from purely virtual processes, i.e. processes where there is insufficient energy for the hidden photon to be produced on-shell:

\begin{itemize}
\item{{\bf Planetary magnetic fields.} The modification to Maxwell's equations governing electrodynamics caused by the kinetic mixing is detectable in the macroscopic, planetary magnetic fields of Earth and Jupiter. The large distance scales involved allow the probing of small hidden photon masses, $m_X = (10^{-15}$--$10^{-12})$~ eV~\cite{Goldhaber:2008xy}, which are irrelevant to the KK scenario, where the minimum mass of a hidden photon mode is 200 $\mu$eV (see Table~\ref{tab:limits} and the corresponding lines in Fig.~\ref{Fig:4d}.)}
\item{{\bf Coulomb potential.} A hidden photon modifies the Coulomb potential in a Yukawa like way~\cite{Okun:1982xi,Popov:1999}, which can be tested precisely in a lab using a Cavendish experiment (measuring the potential difference between concentric spheres). The philosophy is similar to the planetary magnetic field constraints, with the lab length scales resulting in bounds around $m_X = 0.1~\mu$eV.}
\item{{\bf Atomic-spectra.} Transitions between atomic energy levels occur via the emission and absorption of photons. A virtual insertion into this propagator alters the frequency of the transition, changing the atomic spectra \cite{Karshenboim:2010cg,Karshenboim:2010ck,Jaeckel:2010xx}. There is no real production of hidden photons in this case, which leads to some interesting features in the presence of extra dimensions, discussed in Section~\ref{sec:pert}.} 
\item{{\bf Anomalous magnetic moment.} A virtual insertion into the photon propagator means that the anomalous magnetic moment of the electron and muon are sensitive to the hidden photon \cite{Pospelov:2005pr}.}
\item{{\bf Electroweak precision constraints.} The Z-boson mass would receive a contribution from a hidden photon by an insertion into the photon propagator \cite{Hook:2010tw}. Being a virtual correction, this exhibits the same features as the anomalous magnetic moment and atomic spectra. The mass region probed by electroweak precision constraints is eclipsed by the LHC bound (Sec.~\ref{sec:LHC}) in the KK scenario, so it is not investigated further}.
\end{itemize}

The constraints discussed above are summarized in Fig.~\ref{Fig:4d}. It will turn out that the most important bounds on our parameter space are those from the LHC, HB stars and fixed target experiments, with a small additional area ruled out by consideration of the IDPB. All other constraints are effective only in regions of parameter space already ruled out by these few observations and experiments\footnote{The singular exception is the case of SN1987A, which could in principle eclipse the IDPB constraint. However, we think that the existing constraints on hidden photons arising from SN1987A neglect some important plasma effects, so we do not include them here.}.

We also include a discussion of constraints arising from atomic spectra and anomalous magnetic moments, since they give rise to some interesting considerations regarding the cutoff-dependence and perturbativity of the effective theory.
%

It was noted in Section~\ref{sec:model} that the size of extra dimensions are subject to constraints (Table~\ref{tab:limits}). In our scenario, limits on $R$ translate directly to limits on the minimum hidden photon KK mode mass. Fig.~\ref{Fig:4d} shows the minimum masses that are viable for $d = 1 \dots 6$ extra dimensions, overlaid on the constraints in 4 dimensions. This illustrates the reasoning behind the selection of experiments we use to constrain the KK scenario.

\section{Experimental and Observational Constraints}
\label{sec:results}

Let us start by commenting briefly on some generalities regarding the calculation of observational constraints in the presence of a stack of KK modes.
The constraints we present are on the toy model specified in Eq.~\eqref{eq:action}, i.e. the hidden photon has a stack of KK modes, each of which interacts with the same kinetic mixing parameter $\chi_{4}$. The structure of the stack of KK modes depends on the dimension, and therefore different dimensionalities represent different theories.
In this sense constraints on the kinetic mixing should not be compared between models of different dimension.

Processes can now involve a large number of KK modes and we have to sum over all of their contributions.
Typically observables will be dominated by KK modes within a certain mass range. If the spacing between different KK modes, $m_{0}=1/R$, is sufficiently smaller than this range we can approximate the sum over KK modes by an integral,
\begin{equation}
	\sum_{\vec{n}>0} \rightarrow \frac{1}{2^{d}}\int d^{d}k,
\end{equation}
where the prefactor to the integral accounts for it running over the whole space.
This already tells us something about the typical shape that we expect for our constraints.
Let the contribution of a hidden photon of mass $m$ to a given observable be $\chi^{p}\,{\mathcal P}(m)$. Moreover let the dominant contribution
arise from masses between $M_{1}$ and $M_{2}$. Summing over KK modes (requiring $M_{1}\gg m_{0}$) the total contribution then reads,
\begin{equation}
\chi^{p}\frac{2\pi^{\frac{d}{2}}}{\Gamma(\frac{d}{2})}\frac{1}{2^{d}}\int^{M_{2}/m_{0}}_{M_{1}/m_{0}} dk k^{d-1} {\mathcal P}(k\,m_{0})=\chi^{p}\frac{1}{m^{d}_{0}}\frac{2\pi^{\frac{d}{2}}}{\Gamma(\frac{d}{2})}\frac{1}{2^{d}}\int^{M_{2}}_{M_{1}} dM M^{d-1} {\mathcal P}(M).
\end{equation}
The integral on the right hand side is independent of $m_{0}$ and $\chi$. Accordingly we can see that the total contribution scales as
\begin{equation}
\sim \frac{\chi^{p}}{m^{d}_{0}}.
\end{equation}
For small $m_{0}$ any limit arising from a constraint on this observable then strengthens as,
\begin{equation}
\chi\sim m^{-\frac{d}{p}}_{0}
\end{equation}
for decreasing $m_{0}$. Small values of $m_{0}$ are therefore very tightly constrained.

\subsection{Astrophysical Constraints}

\subsubsection{Stellar Lifetime}

The mixing between photon and hidden photon states allows the production of hidden photons in stellar interiors. Once produced, and if not reabsorbed, the hidden states would contribute to the energy loss of the star. If the energy loss through this channel is greater than the visible luminosity, no stellar model can be constructed in which the properties of the Sun (or star) are consistent with observations. Even if the hidden states never leave the Sun, if their presence is sufficiently great, they can contribute unacceptably large levels of non-local energy transfer. The mass and coupling of a hidden photon must be constrained to be consistent with these considerations \cite{Frieman:1987ui,Raffelt:1987yb,Popov:1999,An:2013yfc,Redondo:2013lna}.

For our local star, the Sun, we use the solar model BS05(OP)~\cite{Bahcall:2004pz}. We ignore the effects of the less abundant heavier ions, $^{12}$C, $^{14}$N and $^{16}$O, accounting only for the presence of the lighter elements $^{1}$H, $^{3}$He and $^{4}$He. Since stellar energies are relatively low, $\lesssim$~0.1 MeV, only a negligible fraction of hidden photons that are massive enough to decay to Standard Model particles could be produced, so the only relevant process is the oscillation of photon and hidden photon states. We assume that any photon to hidden photon oscillation is effectively irreversible and contributes to energy loss or unacceptably large non-local energy transfer.

The same stellar energy loss arguments apply to stars elsewhere in the Milky Way, the best studied of which lie on the horizontal branch (HB). There are two significant differences between HB stars and the Sun. The first is the higher luminosity of the HB stars, $L_{HB} \sim 20 L_{\odot}$. The second is their structure. HB stars are globular cluster (i.e. low metallicity) stars and are modelled as having two distinct components: the helium dominated core and the hydrogen dominated shell.

The critical feature of both the Sun and horizontal branch stars is the facility for resonant production of hidden photons at a particular energy, which manifests as the sharp peaks in the constraints shown in Fig.~\ref{Fig:4d}. The HB star actually features two resonance regions (corresponding to the hydrogen shell and helium core structure), with the lower energy resonance being eclipsed by the constraint from the Sun.

These resonances are particularly effective at enhancing constraints in the presence of the Kaluza-Klein stack. When the hidden photon mode spacing is  smaller than the resonance width in the Sun/HB star, at least one mode in the Kaluza-Klein tower will sit in the resonance region. The same scaling is observed as in the other bounds (i.e. the hidden photon effect grows with decreasing mode spacing), but beginning at the peak of the resonance.

Recently, the treatment of \cite{Redondo:2008aa} of solar and horizontal branch constraints on hidden photons has been improved \cite{An:2013yfc,Redondo:2013lna}. The longitudinal hidden photon modes, previously ignored, are actually important in the low energy regime: the contribution from transverse modes to production dies off as $m_{0}^4$, whereas the longitudinal mode contribution dies only as $m_{0}^2$, strengthening the bound into the region probed by light-shining-through-walls experiments. We have included the effect of the longitudinal modes in the constraints for KK towers. The corresponding limits are shown as the yellow and orange regions in Fig.~\ref{Fig:bounds}.

\subsubsection{Intergalactic Diffuse Photon Background}

The intergalactic diffuse photon background (IDPB) could receive a contribution from hidden photons decaying into standard model photons. Naturally the magnitude of the contribution is dependent on the number of hidden photons produced and their survival until the time at which the CMB decouples.

The biggest contribution to the production of a HP with of a given mass arises when the plasma mass in the Universe agrees with the HP mass, i.e. when there is a resonance. Effective thermal production stops at CMB decoupling when the plasma recombines into atoms. Therefore very low mass modes, corresponding to low temperatures, are not produced en masse\footnote{ In fact, there can still be medium-induced resonant effects at lower masses, such as discussed in \cite{Mirizzi:2009iz}, however, they do not lie in a mass region of interest for the LED scenario: the low mass region is already strongly bounded by the cumulative effect of KK modes being resonant at higher energies.}. 
The towers of KK modes (above a certain mass) associated with low $m_0$ hidden photons would still be thermally produced in the earlier Universe and as such the low masses can still be constrained, but a lower cutoff on the modes that contribute to the constraint applies. In any case, data for the diffuse photon spectrum exists only in the energy region of $\sim 10^{4}$~eV and upwards, so only the effect of modes with masses greater than this contribute to the limit. We can, however, still constrain the kinetic mixing of hidden photons with $m_0$ below the range for which data exists by observing the effects of the higher mass modes in the KK tower.

The survival of hidden photons depends upon their decay width. If the decay to electrons and positrons is accessible, i.e. when $m_X > 2 m_e$, this decay mode dominates, eliminating the contribution to the IDBP. However, for $m_X$ below twice the electron mass it is the three-photon decay that dominates. Its rate is given by~\cite{Redondo:2008ec},
\begin{equation}
  \Gamma_{X\to\gamma\gamma\gamma} = \frac{17\alpha^4\chi^2}{11664000\pi^3}\frac{m_X^9}{m_e^8}.
\end{equation}
In order to leave detectable traces in the IDPB, the hidden photons must survive until after the CMB decoupling time, $\sim 10^{12}$~s after the Big Bang. Therefore we enforce the constraint that to be within reach of IDPB exlusion, $\tau = \Gamma_{X\to\gamma\gamma\gamma}^{-1} > 10^{12}$~s. This provides the upper boundary to the purple IDPB region seen for the $d=1,\,2$ cases in Fig.~\ref{Fig:bounds}. For those hidden photons in the parameter range that can survive as dark matter, the same three-photon decay would certainly contribute to the diffuse intergalactic photon spectrum. We use the formulation of~\cite{Redondo:2008ec}, applying the constraint that,
\begin{equation}
	\frac{m_X \tau(m_X,\chi)}{{\rm GeV s}} \lesssim 10^{27} \left(\frac{\omega}{{\rm GeV}}\right)^{1.3}\left(\frac{\Omega_2 h^2(\chi^2)}{0.1}\right),
\label{eq:idpb}
\end{equation}
i.e. that the photon flux from HP decays is not greater than the total photon flux from the IDPB.
In particular, as per~\cite{Redondo:2008ec} (as opposed to~\cite{Yuksel:2007dr}), the hidden photon contribution to the relic density is calculated for each point in parameter space, so the final factor in brackets on the r.h.s. of Eq.~\eqref{eq:idpb} is not unity, but depends on $\chi^2$ (i.e. $\Omega_2 h^2 (\chi^2)$ is not, {\it de facto} equal to 0.1). Solving Eq.~\eqref{eq:idpb} for $\chi$ for each value of $m_0$, including the effects of the tower of KK modes results in the plotted constraint (purple region in Fig.~\ref{Fig:bounds}).

For a given $m_0$, many KK modes contribute to the IDPB signal, with the resulting enhancement spread over a wide spectrum of energies. The experimental analysis (see \cite{Yuksel:2007dr}) separates the spectrum into energy bins--the region in Fig.~\ref{Fig:bounds} represents constraint from the most constraining bin for each value of $m_0$.

\subsection{Constraints from scattering experiments}
\subsubsection{Fixed Target Constraints}

Fixed target, or beam dump, experiments provide very clean testing grounds for new physics, owing in part to the well understood Standard Model backgrounds. Several experiments have previously been used to constrain four dimensional hidden photons~\cite{Bjorken:2009mm}, with some purpose-built experiments underway to further explore the parameter space (see, e.g. Section 3 of~\cite{Essig:2013lka} and references therein). The three fixed target experiments employed here are E137, E141 at Fermilab and E774 at SLAC.

The operating principle of the experiment, in context of a hidden photon search, is that hidden photon bremsstrahlung can propagate through the shielding of the apparatus, before decaying to an electron-positron pair in the vacuum region in front of the detector. The only effect of KK modes on the experiment is to supply additional channels by which the bremsstrahlung can evade the shielded region and we observe a tightening of the constraints at lower masses consistent with this. The slope of the constraint differs from that for the stellar and LHC cases as the kinetic mixing also enters (exponentially) the probability for the hidden photon to decay in the detector region, resulting in a different scaling behavior. It is important to note that this simple strategy applies to searches using event counting. Searches relying on a peak in the invariant mass distribution require a different analysis.

\subsubsection{LHC}
\label{sec:LHC}

LHC constraints on the hidden photon parameter space in the absence of extra dimensions have been computed \cite{Jaeckel:2012yz}, but it is worthwhile to reassess the discovery potential of the LHC in light of our modified phenomenology due to the towers of KK modes.

For the range of kinetic mixing parameters typically of interest, $\chi_{4} < 10^{-3}$ say, we can make a narrow width approximation (NWA) for the hidden photon, so that the width does not affect the cross-section and we can factor the kinetic mixing out of the calculation. If the width is extremely small, once produced on-shell in the beam pipe, the HP may not decay within range of the detectors, but instead escape, leaving only missing transverse energy as a ``signal''. The canonical probe of such a phenomenon is the monojet search.

Alternatively, the hidden photon could decay inside the detector region to two leptons (or else mediate this interaction off-shell), so our candidate events are of dilepton plus jets type. We thus investigate two potential LHC signals of hidden photons: monojets + MET and dileptons.

\subsubsection*{Monojets}

An on-shell hidden photon can be produced in the collision of two protons in s-channel $q g \to \gamma' q$ and and t-channel $q \bar{q} \to \gamma' g$ processes. If such a hidden photon is long lived, i.e. it has a small kinetic mixing parameter $\chi$ that restricts its decays, it can escape the detector region without interacting and contribute to a missing transverse energy (MET) signal, in association with a single jet that balances the (missing) transverse momentum of the hidden photon.

The decay rate of a massive hidden photon into a fermion-antifermion pair is,
\begin{equation}
  \Gamma(X\to f\bar{f}) = \frac{\chi^2 Q_{e/m}^2 \alpha\,m_{X}}{3}
  \sqrt{1-4\frac{m_{f}^2}{m_{X}^2}} \left( 1 + 2 \frac{m_{f}^2}{m_{X}^2}\right).
\end{equation}
The decay time is then $\tau = 1/\Gamma_{\rm tot.}$ where $\Gamma_{\rm tot.}$ must account for all open decay channels (quarks and leptons). We are primarily interested in the LHC for heavy hidden photons, where bounds from other sources are not already strong. Considering for example, a $100$~GeV hidden photon, its total width to charged fermions is $\sim 1 \,\chi^2$~GeV. In order that it's decay length is at least the radius of the ATLAS detector, $\sim 10$~m, we must enforce $\chi \lesssim 5\times 10^{-9}$. This results in a production cross section for a monojet event $\sim 6 \times 10^{-14}$~pb, orders of magnitude too small to give a reasonable event rate at the LHC.

However, due to the mixing with the Z a high mass hidden photon has a branching ratio of $\sim 8$\% to the three species of SM neutrino when all SM decay modes are available (i.e. when $m_X > 2\,m_{\rm top}$) which could provide an enhanced signal. 
Using \texttt{Herwig++} \cite{Bahr:2008pv} and \texttt{CheckMATE} \cite{Drees:2013wra,deFavereau:2013fsa,Cacciari:2011ma,Cacciari:2005hq,Cacciari:2008gp,Read:2002hq,Lester:1999tx,Barr:2003rg,Cheng:2008hk}, we find that for a single hidden photon the additional signal is insufficient to be distinguishable above background uncertainty, even for kinetic mixings of order unity. This may no longer be the case when the effect of all modes in the KK tower is summed, since this provides many more decay modes to neutrinos. However, any possible constraint arising from this is eclipsed by the constraint from the dilepton channel, for which even a single hidden photon provides a constraint~\cite{Jaeckel:2012yz}. As such, monojets do not provide a competitive or novel constraint.

\subsubsection*{Dileptons}

If the hidden photon is produced on-shell and decays into charged leptons, we may search for their tracks. In principle the mediator need not be on-shell, though this provides the dominant contribution to the cross-section, and we assume the resulting dilepton invariant mass distribution to be strongly peaked at the mass of the hidden photon.

We use the result of the ATLAS dilepton search for $Z'$~\cite{ATLAS:2013jma}, mapping the constraints directly to our model, with cross sections generated by \texttt{MadGraph 5} \cite{Alwall:2011uj} from a \texttt{FeynRules} \cite{Alloul:2013bka} model via the UFO format \cite{Degrande:2011ua}. The presence of stacks of KK modes complicates matters slightly. We must factor in the larger contribution to the cross-section due to the presence of multiple KK modes in the same experimental energy bin. For a given mode spacing, $m_0$, the modes are summed per bin. The limit on $\chi$ is calculated for each bin and the tightest constraint arising from this procedure is used as the constraint for a given $m_{X}$.

There is an additional complication arising from multiple modes. In principle, we can sum over the modes in a bin only when they are separated into distinct peaks. A reasonable measure for this is when the mode spacing is greater than the mode width, $\Delta m > \Gamma(m_{X},\chi)$. In the high energy region, $m_{0} \sim$~TeV, the mode width is narrow compared to the mass of the modes. Below this scale, there is a region where the mode width is comparable to the mass of the modes. This is shown as a darkened region of the LHC constraint in Fig.~\ref{Fig:bounds}. Here we cannot ignore interference, which must be calculated at the amplitude level. However, as we will discuss in Section~\ref{sec:pert}, the LHC constraint ends up covering a region of parameter space throughout most of which a perturbative treatment is not valid, rendering interference effects somewhat a moot point.

\subsection{Constraints from low energy precision measurements}
\label{sec:lowenergy}

The previous constraints from astrophysical observations and scattering experiments featured the real production of hidden photons and the associated KK modes. As such, the result is not cutoff dependent, since the maximum energy of the experiment (for example, the temperature in a stellar interior) provided a natural physical upper bound on the masses of KK modes that could be produced. This argument rests upon the hidden photon being produced on-shell, with definite energy and momenta. If the hidden photon were to be produced as an insertion to a propagator, for example in a loop process, the energy available is in principle infinite. This results in a divergence for higher dimensional theories, where the integral over the KK number does not converge. To avoid these divergences, a cutoff must be imposed at some finite energy scale $\Lambda$.

In the constraints of this subsection, there is insufficient energy for real hidden photon production, and we are afflicted by the cutoff dependencies discussed above and in Section~\ref{sec:pert}.

\subsubsection{Atomic Spectral Constraints}

The most stringent of the atomic spectral constraints for four dimensional hidden photons is that associated with the $2s_{1/2} - 2p_{1/2}$ Lamb shift in atomic hydrogen \cite{Karshenboim:2010cg,Karshenboim:2010ck,Jaeckel:2010xx}, so we will limit our attention to this case.
Hidden photons enter the process as a propagator insertion in the Coulomb interaction, which modifies the potential with a Yukawa-like term (in the static limit), $\delta V(r) = -(\chi^2 Z \alpha e^{-m_{X}r})/r$ where $Z$ is the atomic charge, $r$ is the radius and $\alpha$ the fine structure constant. Comparing the modification to the Coulomb potential against the experimental uncertainty in the measurement of the Lamb shift constrains the region plotted in dark red in Fig.~\ref{Fig:bounds}, once the effect of the tower of KK modes is accounted for by summation.

The lack of a natural limit to the hidden photon mass in the propagator means that we must artificially impose a cutoff in the sum. In Fig.~\ref{Fig:bounds}, a finite UV cutoff is used, corresponding to the point at which the mixing parameter $\chi$ becomes non-perturbative due to the cumulative effect of many KK modes. See Section \ref{sec:pert} for a detailed discussion of cutoff dependency and the related problem of perturbativity.

\subsubsection{Electron and Muon $g-2$ Constraints}

\begin{figure}[]
\centerline{
\subfigure[$d=1$]{\includegraphics[width=0.5\textwidth]{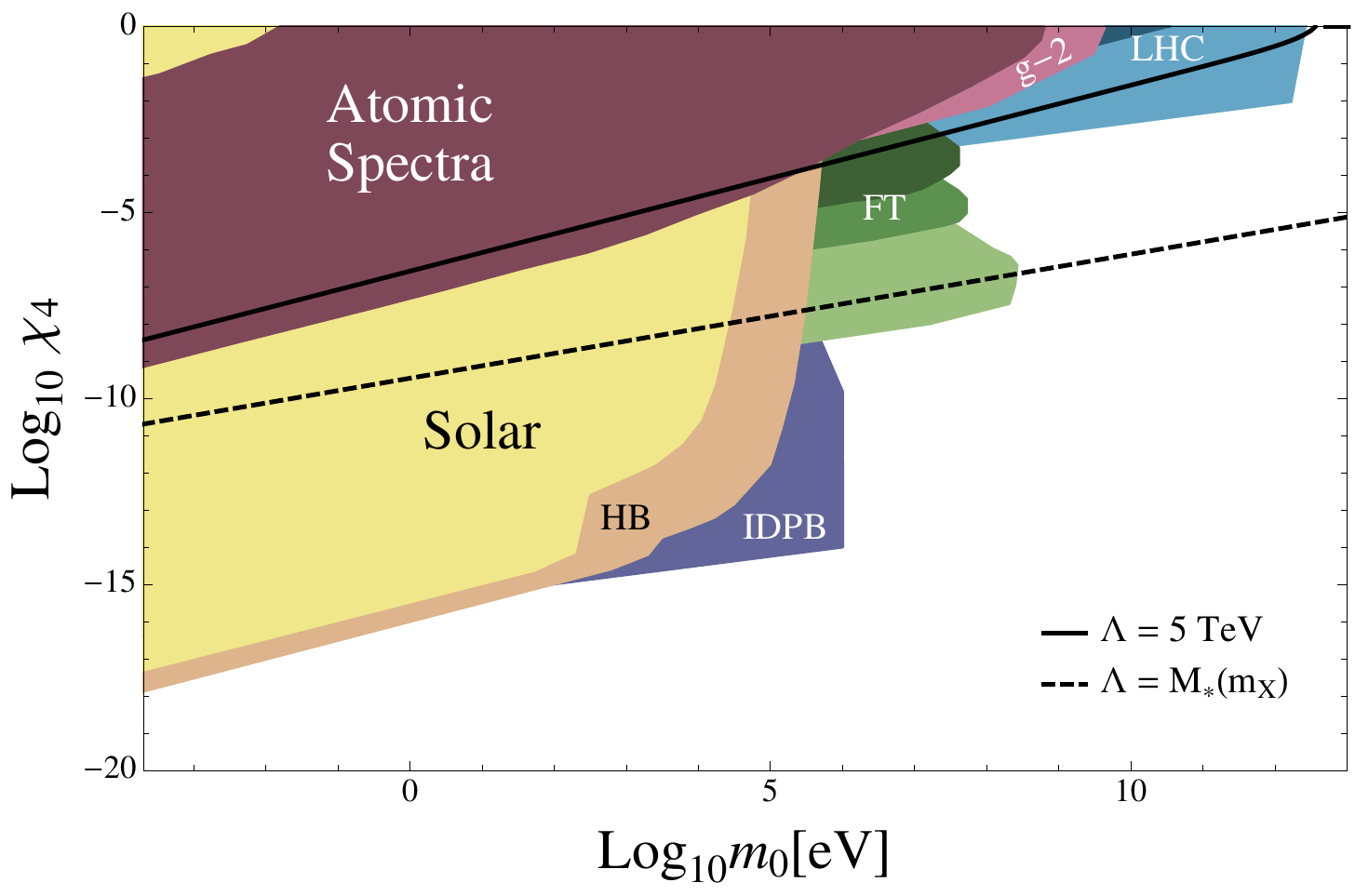}}
\subfigure[$d=2$]{\includegraphics[width=0.5\textwidth]{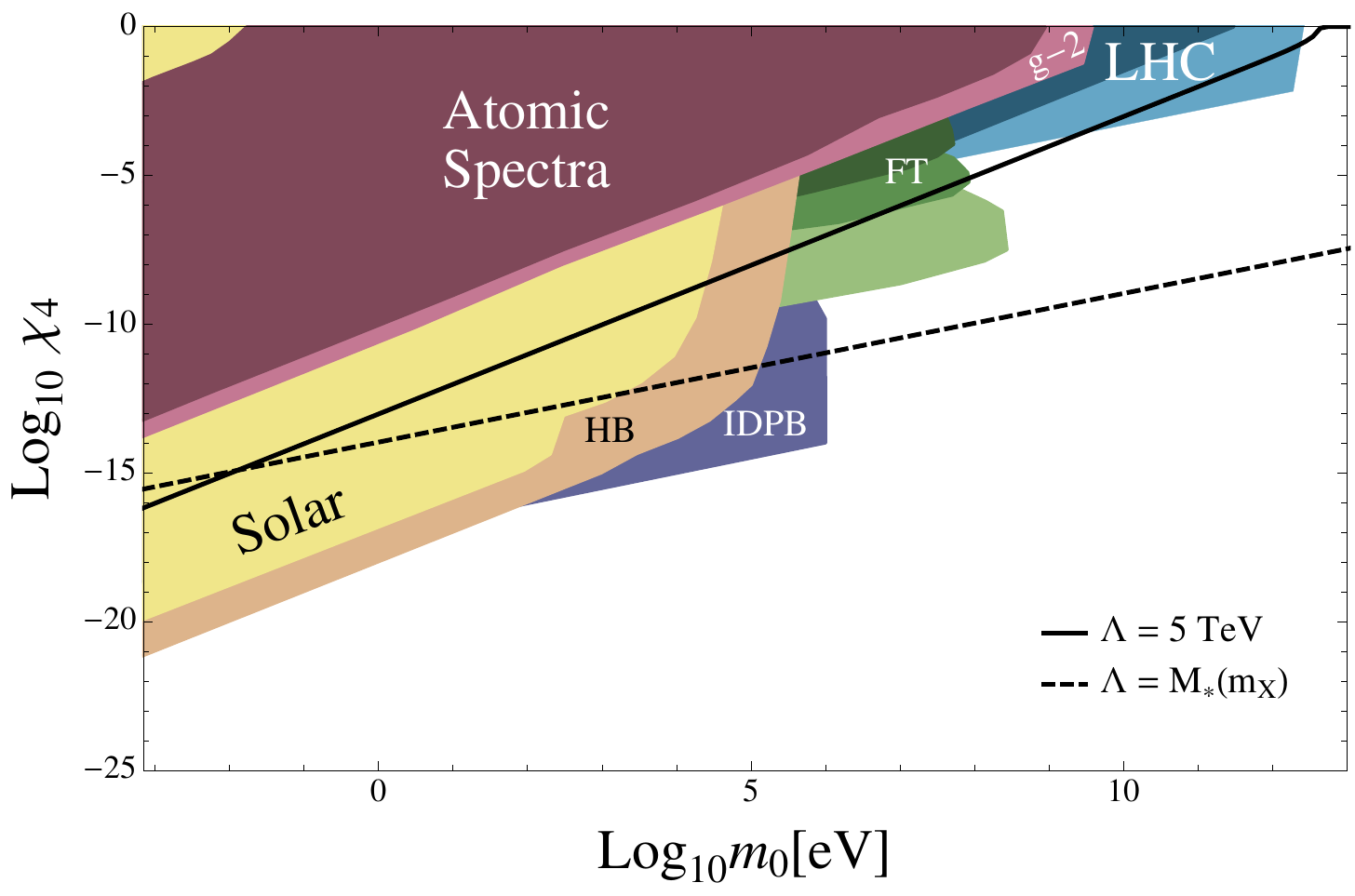}}}

\centerline{
\subfigure[$d=3$]{\includegraphics[width=0.5\textwidth]{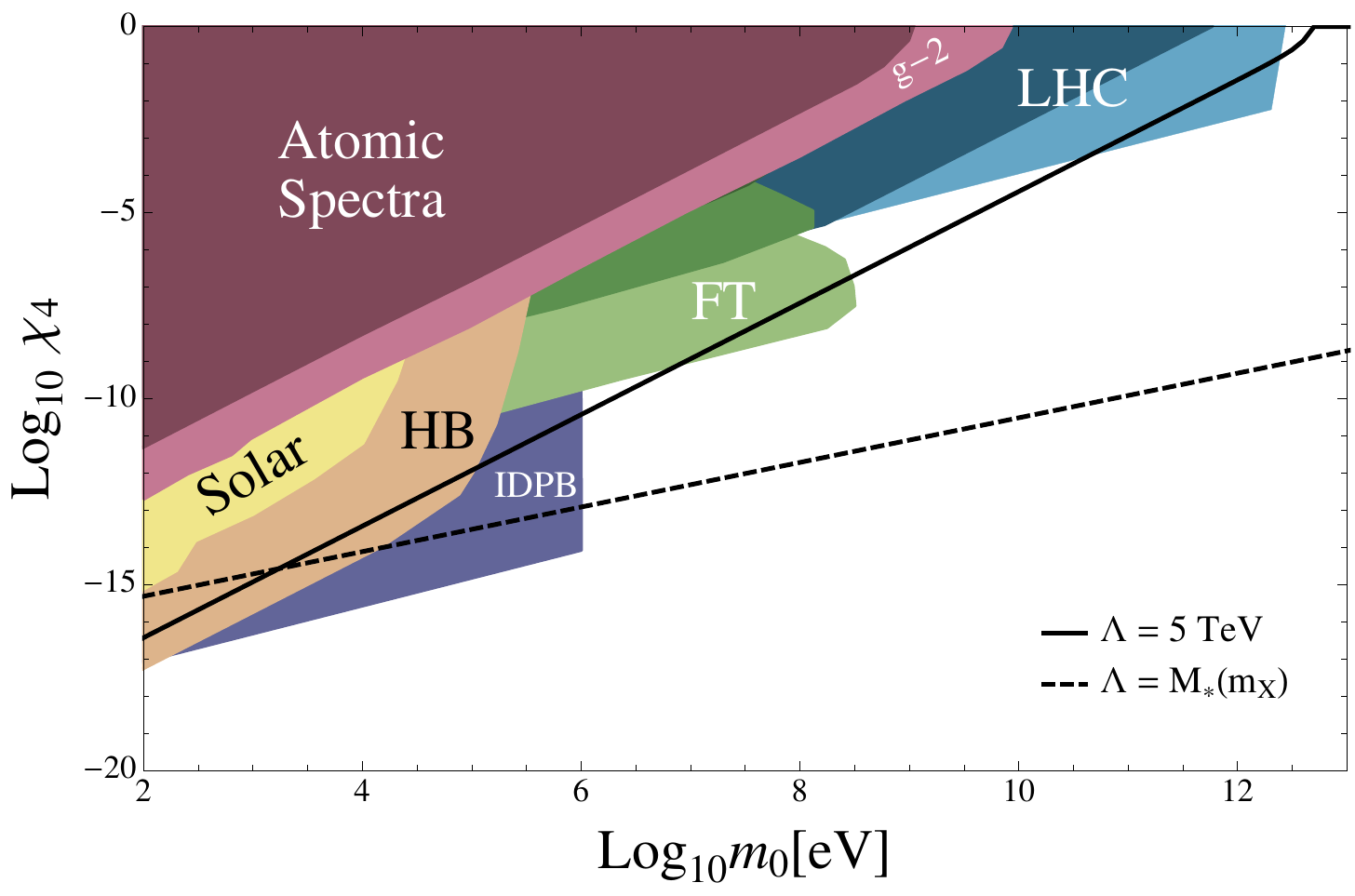}}
\subfigure[$d=4$]{\includegraphics[width=0.5\textwidth]{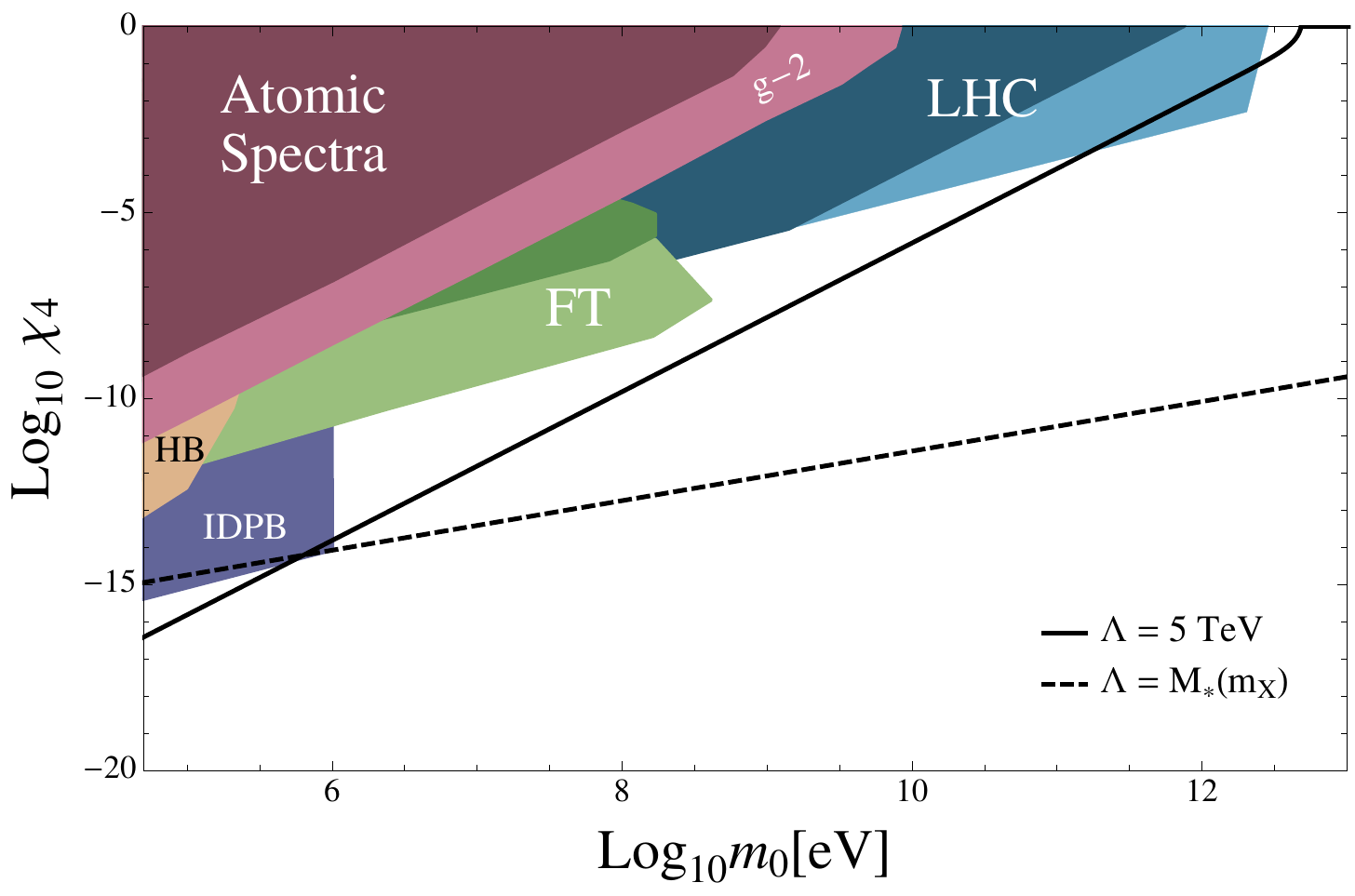}}}

\centerline{
\subfigure[$d=5$]{\includegraphics[width=0.5\textwidth]{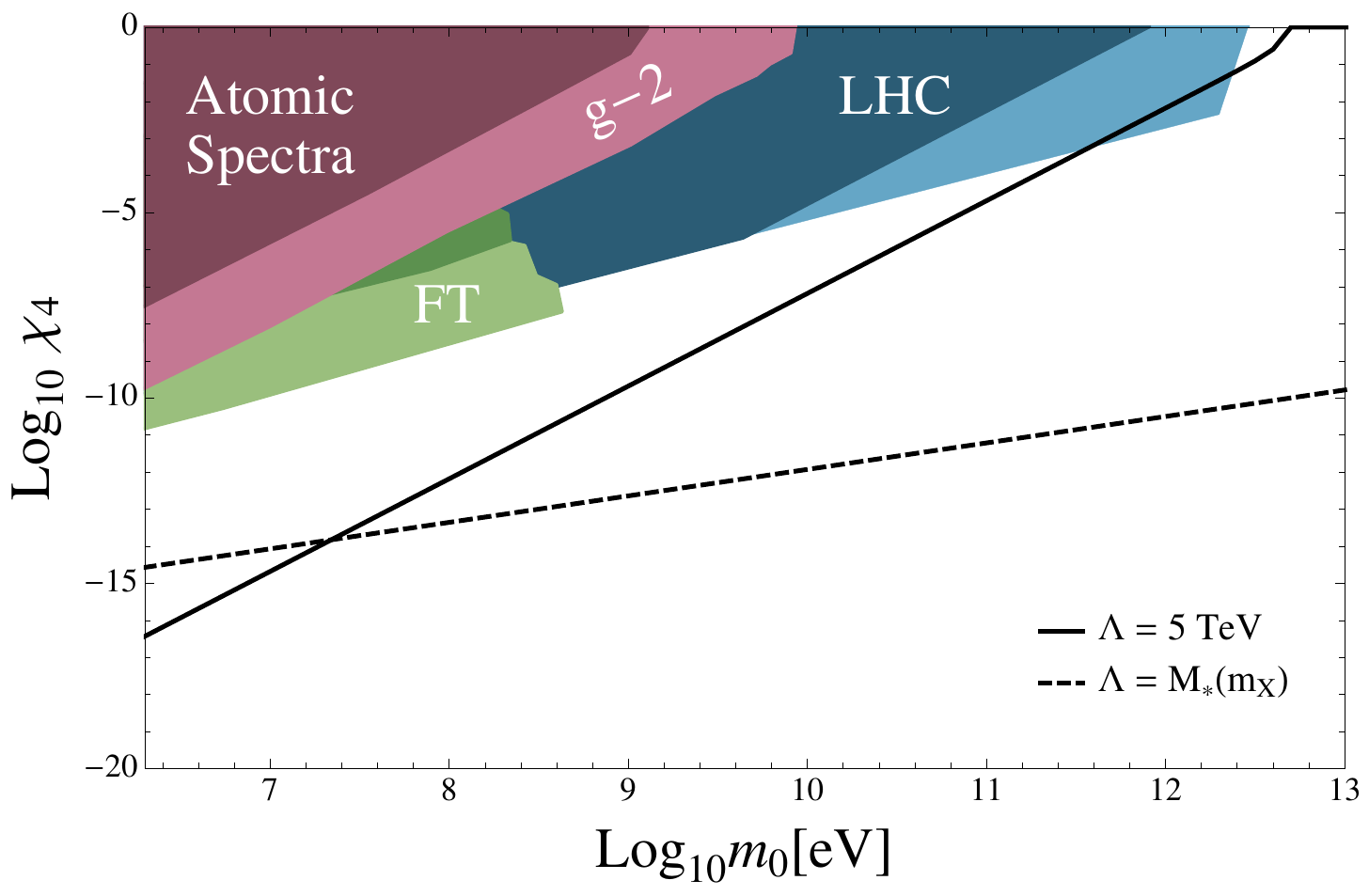}}
\subfigure[$d=6$]{\includegraphics[width=0.5\textwidth]{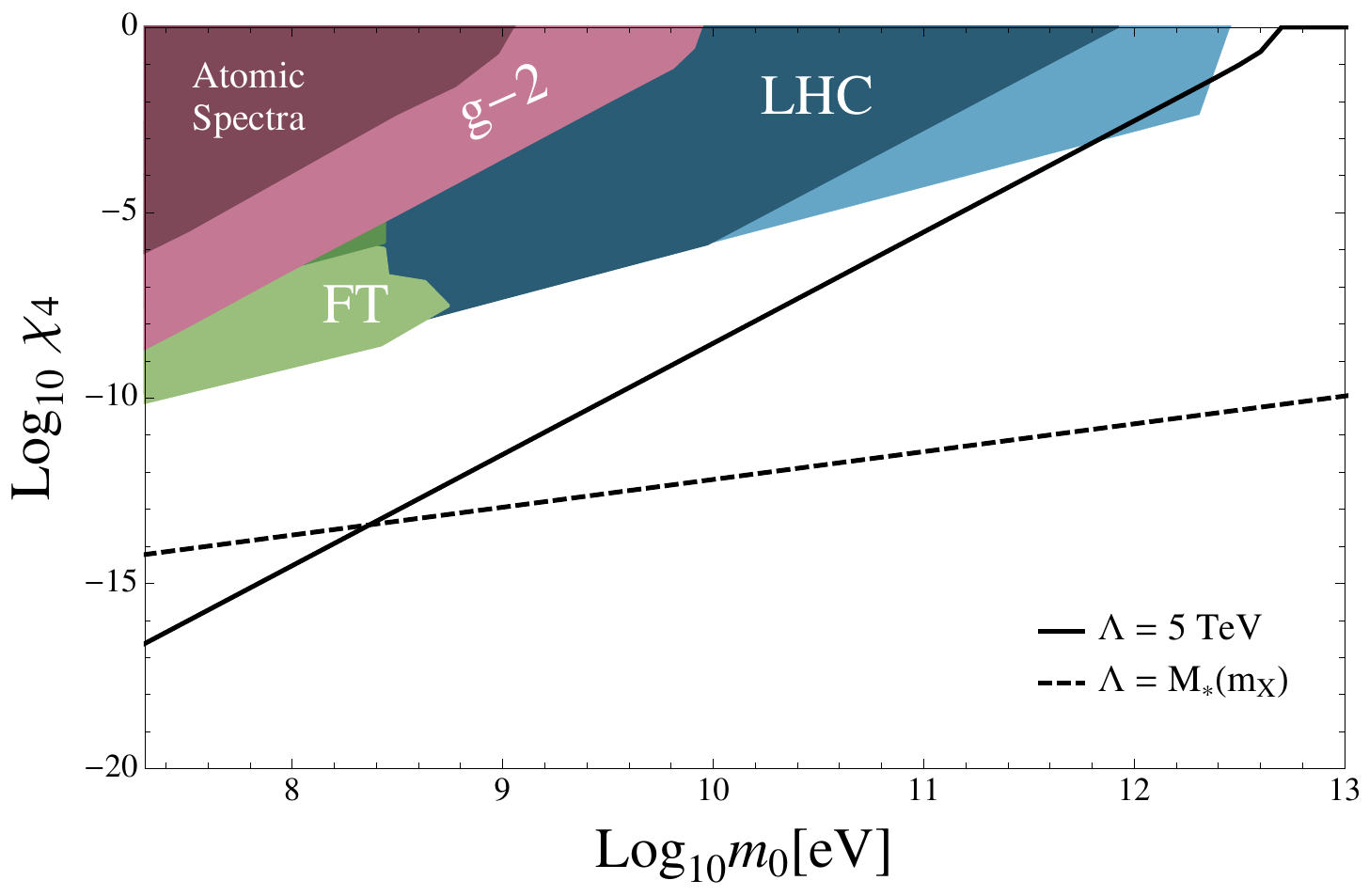}}}

\caption[]{Limits on hidden photon parameter space for $d=1,\dots,6$ extra dimensions. The mass constrained, $m_0$, represents the lowest KK mode mass, i.e. also the spacing between modes. The lowest value of the mass axis is set by constraints on the maximum size, $R$, of the extra dimensions from graviton searches, see Table~\ref{tab:limits}. The black lines indicate regions (above) where a perturbative treatment fails if a cut off at the indicated scale is employed. Refer to the text for further discussion.}\label{Fig:bounds}
\end{figure}

The electron and muon $g-2$ are among the most precisely measured of all physical observables. The anomalous magnetic moment, $a_{e (\mu)} = \frac{g_{e (\mu)}-2}{2}$, arises from the electron (muon) vertex correction and would be sensitive to hidden photon insertions to the photon propagator \cite{Pospelov:2005pr}.

The analysis of magnetic dipole moment constraints proceeds very much analogously to the constraints from atomic spectra, and we apply the same cutoff scheme to the results shown in Fig.~\ref{Fig:bounds}. There is an added feature in the case of magnetic dipole moments that there are two measurements which need to be combined to give a constraint\footnote{One can also use this method to combine measurements of exotic atomic spectra, but, due to the experimental precision in the measurements, the Lamb shift alone imposes a stronger constraint.}. Both the electron and the muon magnetic moments receive a contribution proportional to $\chi^2$. However, neither measurement alone provides a constraint since one could adjust $\alpha$ to cancel out the HP contribution. Combining the measurements with the renormalisation procedure described in \cite{Jaeckel:2010xx} resolves this ambiguity and also results in the appropriate decoupling behavior in the low mass region. Note, that after summation over KK modes, the bound (pink region in Fig.~\ref{Fig:bounds}) strengthens as the mass decreases in the LED scenario.

\subsection{Summary of constraints}

The combined constraints on the ($m_0$,$\chi$) parameter space are shown for each $d=1,\dots,6$ extra dimensions in Fig.~\ref{Fig:bounds}. The horizontal axis corresponds to $m_0$, the minimum KK mode mass and mass spacing. For each point in the parameter space the effect of all higher mass modes has been included, with the mode spacing set by the mass at the point in the parameter space being probed. The lowest mass in each plot corresponds to the limit from Table~\ref{tab:limits}. There is in principle no upper limit to the parameter space, and indeed the KK stack continues up to an arbitrarily high mass scale. However, no known phenomena operate with sufficiently high energy to produce the higher mass modes, so we can derive no constraints.

\section{Limits of the effective theory, perturbativity, etc.}
\label{sec:pert}

When calculating the amplitude for a process involving the hidden photon, we must sum over the Kaluza-Klein modes with masses low enough that they can be produced in the collider, stellar interior or whatever relevant factory. For situations where the hidden photon is produced on-shell, there is a natural cutoff, $\Lambda = E_{\rm max.}$, for the maximum mode mass at the energy of the experiment. (We discuss cases where the hidden photon is produced off-shell below).

\begin{figure}[]
\centerline{\subfigure[$d=1$]{\includegraphics[width=0.5\textwidth]{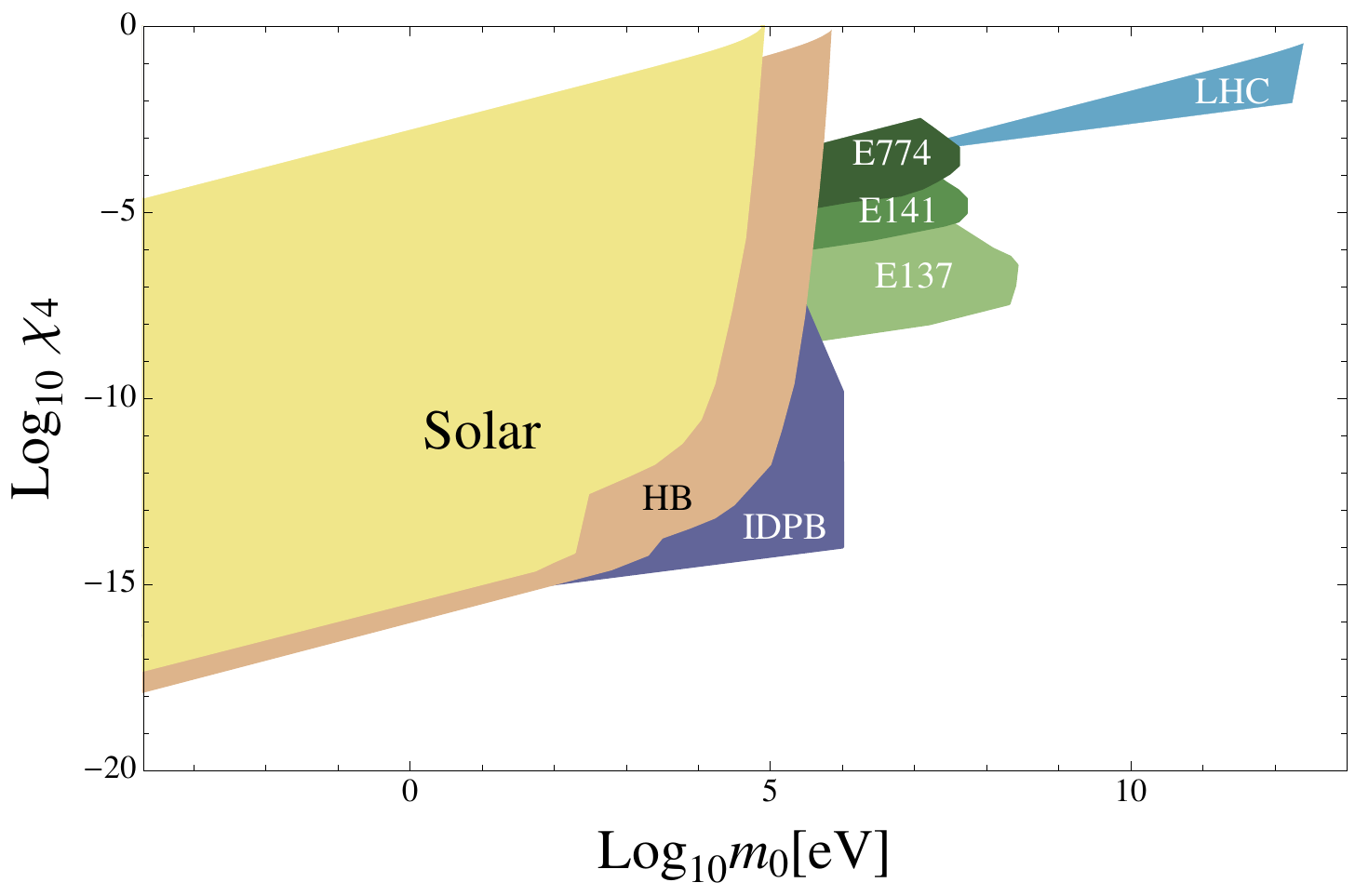}}
\subfigure[$d=4$]{\includegraphics[width=0.5\textwidth]{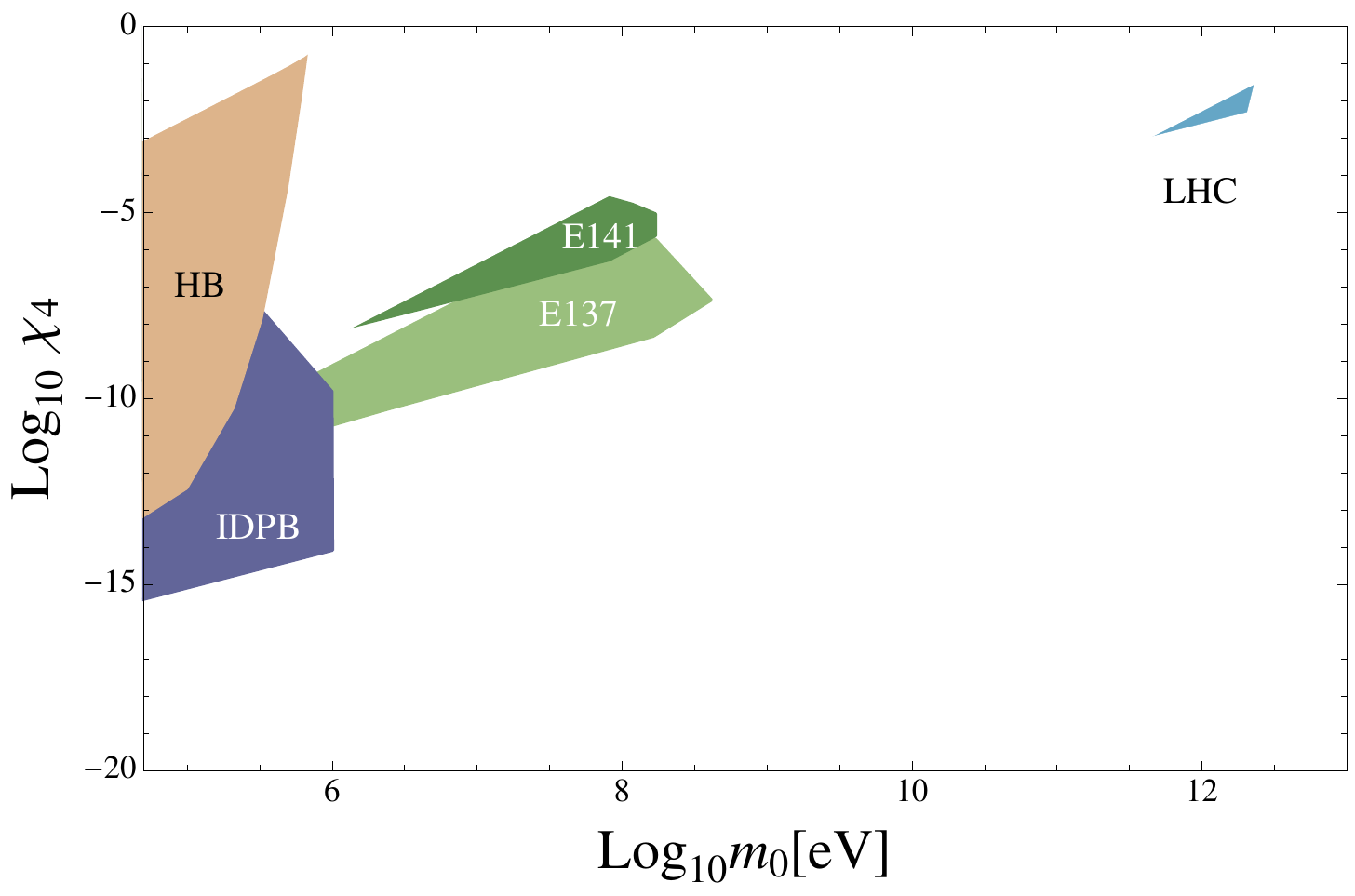}}}
\caption[]{Limits on hidden photon parameter space for $d=1,4$ extra dimension. Perturbativity limits have been enforced on a bound-by-bound basis, with modes included up to the energy scale of each constraint. Regions where a perturbative treatment is invalid are not shown as constrained (see text for details).}\label{Fig:pertbounds}
\end{figure}

The higher the cutoff scale, the more Kaluza-Klein modes contribute to a process. The stack of modes all mediate the same force, so each available mode increases an effective coupling (in this case, the kinetic mixing parameter) of the theory. When the effective coupling becomes large, the theory is no-longer weakly coupled and becomes non-perturbative. We can define a region of validity of our perturbative treatment as shown, by enforcing,
\begin{equation}
  \chi_{\rm pert.}^2 = \chi^2 \times \int_1^{\frac{\Lambda}{m_0}}d^{\,d} k
  =\chi^2 \times \int_1^{\frac{\Lambda}{m_0}}\frac{2\pi^{\frac{d}{2}}}{\Gamma(\frac{d}{2})} k^{d-1} dk
  \ll 1.
\label{eq:pert}
\end{equation}
Regions of parameter space outside of this domain are not \textit{de facto} excluded. It is simply that for some combinations of parameter values (lower masses and higher mixings), the theory is so strongly coupled that our perturbative treatment is not valid, and we cannot make a statement about compatibility of the theory with experiment. These non-perturbative regions are whited-out in Fig.~\ref{Fig:pertbounds}.

Amplitudes involving virtual hidden photons do not present us with such a ``natural'' maximum energy. The 4-momentum of the virtual HP is integrated over, in principle to infinity. If the amplitude for a process does not decouple sufficiently quickly at high energies, which is the case for many processes in which the hidden photon is off-shell, it can diverge with the log or some power of the energy. To tame this divergence, we must insert by hand a UV cutoff.
Any finite cutoff defines a region in which the theory is non-perturbative. We will consider three kinds of cutoff scheme. 

Perhaps the simplest scheme is to employ a fixed cutoff at some suitable scale. In Fig.~\ref{Fig:bounds} we show the line in parameter space that defines the lower boundary of the non-perturbative region for a cutoff at $\Lambda = 5$~TeV (i.e. a little greater than the current minimum scale of extra dimensions from graviton searches in Table~\ref{tab:limits}). This seems to rule out a perturbative treatment of almost all of the experimentally accessible parameter space, but these are very strict requirements to enforce. While it is certainly reasonable to impose this condition on the LHC constraints, it is not strictly necessary to require the theory to be perturbative anywhere above the experimental energy at which it is being probed for our treatment to be valid. Consider the fixed target constraints. In order for our treatment of those to be valid, we must only require perturbativity up to the $\sim$GeV scale. This significantly raises the boundary at which perturbativity breaks down.

Another possible choice for $\Lambda$ is extra dimensional Planck scale $M_{\star}$. Using the relationship $M_{\star}^{2+d} = M_{\rm Pl}^2/V_{d}$, where $V_{d}$ is the extra dimensional volume, $(2 \pi R)^{d}$, we can see that the cutoff is proportional to the inverse size of the extra dimension to some power. We can phrase this in terms of the lowest KK mode mass as
\begin{equation}
  M_{\star} = \left(\frac{M_{\rm Pl}^2}{(2\pi)^d} m_0^d\right)^{\frac{1}{2+d}}.
\end{equation}
The application of a cutoff represents summing over only a finite number of modes equal to $\Lambda/m_0$. If we choose $\Lambda = M_{\star}$ as a cutoff, then both $\Lambda$ and the number of modes summed over vary with the mode mass, so the cutoff changes for different points in parameter space. As indicated by the dashed line in Fig.~\ref{Fig:bounds}, this cutoff rules out a perturbative treatment of essentially all the accessible parameter space for $d \ge 4$, and large sections for $d= 1\dots 3$. However, we are working in a low energy regime, using an effective theory that is understood to be a reasonable description only up to some finite energy scale, which could be lower than $M_{\star}$.
%

The second scheme we consider is to cut off the integration at the point at which the effective kinetic mixing parameter becomes large enough that a perturbative treatment is no longer valid. The constraints from atomic spectra and $g-2$ bounds in Fig.~\ref{Fig:bounds} are shown as they arise when only KK modes up to the $\chi_{\rm eff.} = 1$ limit are included. The effect of this cutoff is only significant in the $d=1,2$ cases, leaving a small corner of parameter space unconstrained by the either $g-2$ or atomic spectra bounds, but in any case ruled out by solar lifetime (and HB star) bounds. If additional KK modes are included (up to some fixed high scale, say 5 TeV) the upper left corner of parameter space is also excluded by these experiments, but one should not trust a perturbative treatment in such a regime, where the kinetic mixing parameter is non-perturbatively large.

The final cutoff scheme to consider is to cut off the effect of KK modes above the relevant experimental energy. This is equivalent to a fixed cutoff, with the cutoff changing between constraints from different sources. To trust the constraint from horizontal branch stars, for example, we require that the effective kinetic mixing remain small only when summing over modes with masses less than approximately an MeV. Imposing the fixed cutoff at the energy scale of each experiment independently and then whiting out the regions that the cutoff renders non-perturbative generates the plots shown for $d=1,4$ extra dimensions in Fig.~\ref{Fig:pertbounds}.

\section{Conclusions}
\label{sec:conc}

In UV completions of the Standard Model, in particular string theory, both additional U(1) gauge groups and extra dimensions naturally emerge. Here, we have employed a low energy effective theory for a toy model in order to capture something of the flavour of the phenomenology of such a setup.

The salient feature of the bounds is increasingly tight constraints as the mass of the lightest hidden photon KK mode (i.e. the mode spacing) decreases. This owes to the increasing number of Kaluza-Klein modes available to any given production mechanism as the modes become lighter and more tightly spaced.

We found that, true to the non-renormalizable nature of extra dimensional theories, constraints arising from processes where the hidden photon was not produced on-shell suffered from cutoff dependence. Fixing a cutoff defines a region of the parameter space where the theory is non-perturbative and a perturbative treatment is not sufficient to make any statement about the experimental validity of the theory. The non-perturbative regions can be large and limit our understanding of the behaviour of such theories.

It is clearly unrealistic to draw conclusions about the truth of any given UV completion from such a crude model, but our investigation does highlight that the constraints on the hidden photon mass--kinetic mixing plane do strongly depend on the mechanism of mass generation and that they can also depend on the additional states that a UV completion may provide.

\section*{Acknowledgements}

We would like to thank thank Javier Redondo for useful discussions. CJW thanks ITP Heidelberg for generous hospitality while a large portion of this work was carried out.


\begin{footnotesize}

\end{footnotesize}


\end{document}